\documentclass{aa}  

\usepackage{graphicx}
\usepackage{txfonts}
\usepackage{xcolor}
\usepackage{lastpage}
\usepackage{orcidlink}
\usepackage{mathtools}

\newcommand{\oiii}{[O\,\textsc{iii}]}
\newcommand{\nii}{[N\,\textsc{ii}]}
\newcommand{\sii}{[S\,\textsc{ii}]}

\newcommand{\siii}{[S\,\textsc{iii}]}
\newcommand{\oii}{[O\,\textsc{ii}]}

\newcommand{\hii}{H\,\textsc{ii}}
\newcommand{\ha}{H$\alpha$}
\newcommand{\hb}{H$\beta$}

\newcommand{\te}{$T_{\rm e}$}

\newcommand{\kms}{km~s$^{-1}$}

\newcommand{\HD}{\label{HD} Astronomisches Rechen-Institut, Zentrum f\"{u}r Astronomie der Universit\"{a}t Heidelberg, M\"{o}nchhofstra\ss e 12-14, D-69120 Heidelberg, Germany}
\newcommand{\UT}{\label{UT} McDonald Observatory, The University of Texas at Austin, 1 University Station, Austin, TX 78712-0259, USA}
\newcommand{\Carnegie}{\label{Carnegie} Observatories of the Carnegie Institution for Science, 813 Santa Barbara Street, Pasadena, CA 91101, USA}
\newcommand{\UChile}{\label{UChile} Departamento de Astronom\'{i}a, Universidad de Chile, Camino del Observatorio 1515, Las Condes, Santiago, Chile}
\newcommand{\UNAMCU}{\label{UNAMCU} Universidad Nacional Aut\'onoma de M\'exico, Instituto de Astronom\'ia, AP 70-264, CDMX 04510, M\'exico}
\newcommand{\APO}{\label{APO} Apache Point Observatory and New Mexico State University, P.O.\ Box 59, Sunspot, NM 88349-0059, USA}    
\newcommand{\NYUAD}{\label{NYUAD} New York University Abu Dhabi, P.O. Box 129188, Abu Dhabi, UAE}
\newcommand{\Utah}{\label{Utah} Department of Physics and Astronomy, University of Utah, 115 S. 1400 E., Salt LakeCity, UT 84112, USA}   
\newcommand{\UCatolica}{\label{UCatolica} Instituto de Astronom\'ia, Universidad Cat\'olica del Norte, Av. Angamos 0610, Antofagasta, Chile}
\newcommand{\UDP}{\label{UDP} Instituto de Estudios Astrof\'isicos, Facultad de Ingenier\'ia y Ciencias, Universidad DiegoPortales, Av. Ej\'ercito Libertador 441, Santiago, Chile}
\newcommand{\CITA}{\label{CITA} Canadian Institute for Theoretical Astrophysics (CITA), University of Toronto, 60 St George St, Toronto, ON M5S 3H8, Canada}
\newcommand{\MPIA}{\label{MPIA} Max-Planck-Institut f\"{u}r Astronomie, K\"{o}nigstuhl 17, D-69117, Heidelberg, Germany}
\newcommand{\STScI}{\label{STScI} Space Telescope Science Institute, 3700 San Martin Drive, Baltimore, MD 21218, USA}
\newcommand{\UNAME}{\label{UNAME} Universidad Nacional Aut\'onoma de M\'exico, Instituto de Astronom\'ia, AP 106, Ensenada22800, BC, M\'exico}
\newcommand{\IAC}{\label{IAC} Instituto de Astrof\'\i sica de Canarias, La Laguna, Tenerife, E-38200, Spain}
\newcommand{\Firenze}{\label{Firenze} Dipartimento di Fisica e Astronomia, Università degli Studi di Firenze, Via G. Sansone 1, 50019, Sesto F.no (Firenze), Italy}
\newcommand{\UCNA}{\label{UCNA} Universidad Cat\'olica del Norte, N\'ucleo UCN en Arqueolog\'ia Gal\'actica - Inst. de Astronom\'ia, Av. Angamos 0610, Antofagasta, Chile}
\newcommand{\UCNB}{\label{UCNB} Universidad Cat\'olica del Norte, Departamento de Ingenier\'ia de Sistemas y Computaci\'on, Av. Angamos 0610, Antofagasta, Chile}
\newcommand{\NYUCASS}{\label{NYUCASS} Center for Astrophysics and Space Science (CASS), New York University Abu Dhabi, P.O. Box 129188, Abu Dhabi, UAE}
\newcommand{\LaSerena}{\label{LaSerena} Department of Astronomy, Universidad de La Serena, Av. Raul Bitran \#1302, La Serena, Chile}
\newcommand{\NAOC}{\label{NAOC} Chinese Academy of Sciences South America Center for Astronomy, National Astronomical Observatories, CAS, Beijing 100101, China}  
\newcommand{\SAIMSU}{\label{SAIMSU} Sternberg Astronomical Institute, Moscow State University, Moscow, 119234, Russia}

\usepackage{color}
\usepackage{natbib,twoopt}
\usepackage[hyphenbreaks]{breakurl}
\usepackage{hyperref}      %% to avoid \citeads line fills, add "draft"  [breaklinks]
\bibpunct{(}{)}{;}{a}{}{,}             %% natbib format for A&A and ApJ
\hypersetup{colorlinks=true,linkcolor=blue,citecolor=blue,filecolor=blue,urlcolor=blue}
\makeatletter
  \newcommandtwoopt{\citeads}[3][][]{\href{http://adsabs.harvard.edu/abs/#3}%
    {\def\hyper@linkstart##1##2{}%
     \let\hyper@linkend\@empty\citealp[#1][#2]{#3}}}
  \newcommandtwoopt{\citepads}[3][][]{\href{http://adsabs.harvard.edu/abs/#3}%
    {\def\hyper@linkstart##1##2{}%
     \let\hyper@linkend\@empty\citep[#1][#2]{#3}}}
  \newcommandtwoopt{\citetads}[3][][]{\href{http://adsabs.harvard.edu/abs/#3}%
    {\def\hyper@linkstart##1##2{}%
     \let\hyper@linkend\@empty\citet[#1][#2]{#3}}}
  \newcommandtwoopt{\citeyearads}[3][][]%
    {\href{http://adsabs.harvard.edu/abs/#3}
    {\def\hyper@linkstart##1##2{}%
     \let\hyper@linkend\@empty\citeyear[#1][#2]{#3}}}
\makeatother
\begin{document}

   \title{
   SDSS-V LVM: Verifying what, and where, \\ the `Galactic Center' Lobe is 
   }

\author{
       K. Kreckel\inst{\ref{HD}} \thanks{\email{kathryn.kreckel@uni-heidelberg.de}}  \and % $^\orcid{0000-0001-6551-3091}$
       O. V. Egorov\inst{\ref{HD}}  \and % $^\orcid{0000-0002-4755-118X}$
       N. Drory\inst{\ref{UT}} \and 
       G. A. Blanc\inst{\ref{Carnegie},\ref{UChile}} \and
       J. E. M\'endez-Delgado\inst{\ref{UNAMCU}} \and %$^\orcid{0000-0002-6972-6411}$
       S. Kabanovic\inst{\ref{HD}} \and
       D. Bizyaev\inst{\ref{APO},\ref{SAIMSU}} \and
       J. R. Brownstein\inst{\ref{Utah}} \and
       E. Egorova\inst{\ref{HD}} \and
       J. G. Fern\'andez-Trincado\inst{\ref{UCNA},\ref{UCNB}} \and 
       P. García\inst{\ref{NAOC},\ref{UCatolica}}\orcidlink{0000-0002-8586-6721} \and
       J. D. Gelfand\inst{\ref{NYUAD},\ref{NYUCASS}} \and
       E. J. Johnston\inst{\ref{UDP}} \and
       I. Katkov\inst{\ref{NYUAD},\ref{NYUCASS},\ref{SAIMSU}} \and
       J. Kollmeier\inst{\ref{Carnegie}, \ref{CITA}} \and
       F.-H. Liang\inst{\ref{HD}}\orcidlink{0000-0003-2496-1247} \and
       K. S. Long\inst{\ref{STScI}} \and
       A. Z. Lugo-Aranda\inst{\ref{UNAME}} \and
       A. Meija\inst{\ref{Carnegie}} \and
       H.-W. Rix\inst{\ref{MPIA}} \and
       A. Roman-Lopes\inst{\ref{LaSerena}} \and
       C. G. Rom\'an-Z\'u\~niga\inst{\ref{UNAME}} \and
       N. Sattler\inst{\ref{HD}} \and
       S. F. Sanchez\inst{\ref{UNAME}, \ref{IAC}} \and
       E. Zari\inst{\ref{Firenze}} \and
       R. de J. Zerme\~no\inst{\ref{UNAMCU}}            }

\institute{\tiny
\HD \and  
\UT \and
\Carnegie \and
\UChile \and
\UNAMCU \and
\APO \and
\SAIMSU \and
\Utah \and
\UCNA \and
\UCNB \and
\NAOC \and
\UCatolica \and
\NYUAD \and
\NYUCASS \and
\UDP \and
\CITA \and
\STScI \and
\UNAME \and
\MPIA \and
\LaSerena \and
\IAC \and
\Firenze 
}

   \date{Received XX; accepted XX}

% \abstract{}{}{}{}{} 
% 5 {} token are mandatory
 
  \abstract
  {
The so-called `Galactic Center' Lobe (GCL) is an extended ($\sim 1^\circ$) radio continuum feature situated above the Galactic Plane, for which the literature contains varying claims about both its nature and location. 
Using new optical integral field spectroscopic observations from the SDSS-V Local Volume Mapper, we confirm the characterization of the GCL as a foreground photoionized \hii\ region, not associated with the Galactic center. We present a new analysis of the ionized gas morphology, line ratio diagnostics, and kinematics.
From our \siii$\lambda$9532 emission line map, which suffers the least extinction, we identify ionized gas emission throughout a closed outer loop, which does not fill the GCL interior. All optical line ratio diagnostics are consistent with photoionization. By comparing the ionized gas reddening from the Balmer decrement with 3D dust maps, we directly constrain the distance to the GCL to $\sim$2~kpc. \nii$\lambda$6583 line kinematics show a uniform velocity structure across the GCL, further confirming that the entire bubble is one structure. 
The size and emission line morphology is strongly reminiscent of that seen in the nearby Barnard's Loop, providing a possible analog to explain how this outer shell may be photoionized by a more distant and off-center embedded young cluster. 
We suggest the acronym GCL be repurposed to instead abbreviate the name `Greatly Confused Loop’. 
}
   \keywords{ISM: general --
                HII regions --
                Galaxy: local insterstellar matter --
                ISM: clouds
               }

   \maketitle

\section{Introduction}

The Galactic Center of our Milky Way is a formidable engine of astrophysical activity, characterized by extreme molecular densities, strong magnetic fields, and a star formation rate per unit volume that exceeds the Galactic disk average by at least one order of magnitude \citep{Morris1996, Henshaw2023ASPC..534...83H}. Understanding the injection of energy and momentum from this central region out to larger scales can be achieved by mapping the vertical structures, tracing how energy is channeled from the nucleus into the Galactic halo. Dominating the radio continuum morphology of this region is the \textit{Galactic Center Lobe} (GCL), a towering loop-like structure extending nearly $1^{\circ}$ above the Galactic plane. Since its identification by \citet{sofue1984}, the GCL has served as a Rorschach test for Galactic astrophysics, and depending on the wavelength of observation it has been interpreted variously as a magnetic flux tube \citep{Heyvaerts1988ApJ...330..718H, Uchida1985Natur.317..699U}, a local supernova remnant \citep{Uchida1994ApJ...421..505U},  a galactic wind \citep{Bland-Hawthorn2003, Law2010ApJ...708..474L}, an unknown energetic event \citep{Heywood2019Natur.573..235H}, or a foreground \hii\ region \citep{Tsuboi2020PASJ...72L..10T, Anderson2024ApJ...969...43A}. 
This has led to a forty-year struggle to separate genuine nuclear features from the foreground Galactic disk, and it remains unclear in the literature what and where this object is.

\begin{figure*}
    \centering
    \includegraphics[width=0.8\linewidth]{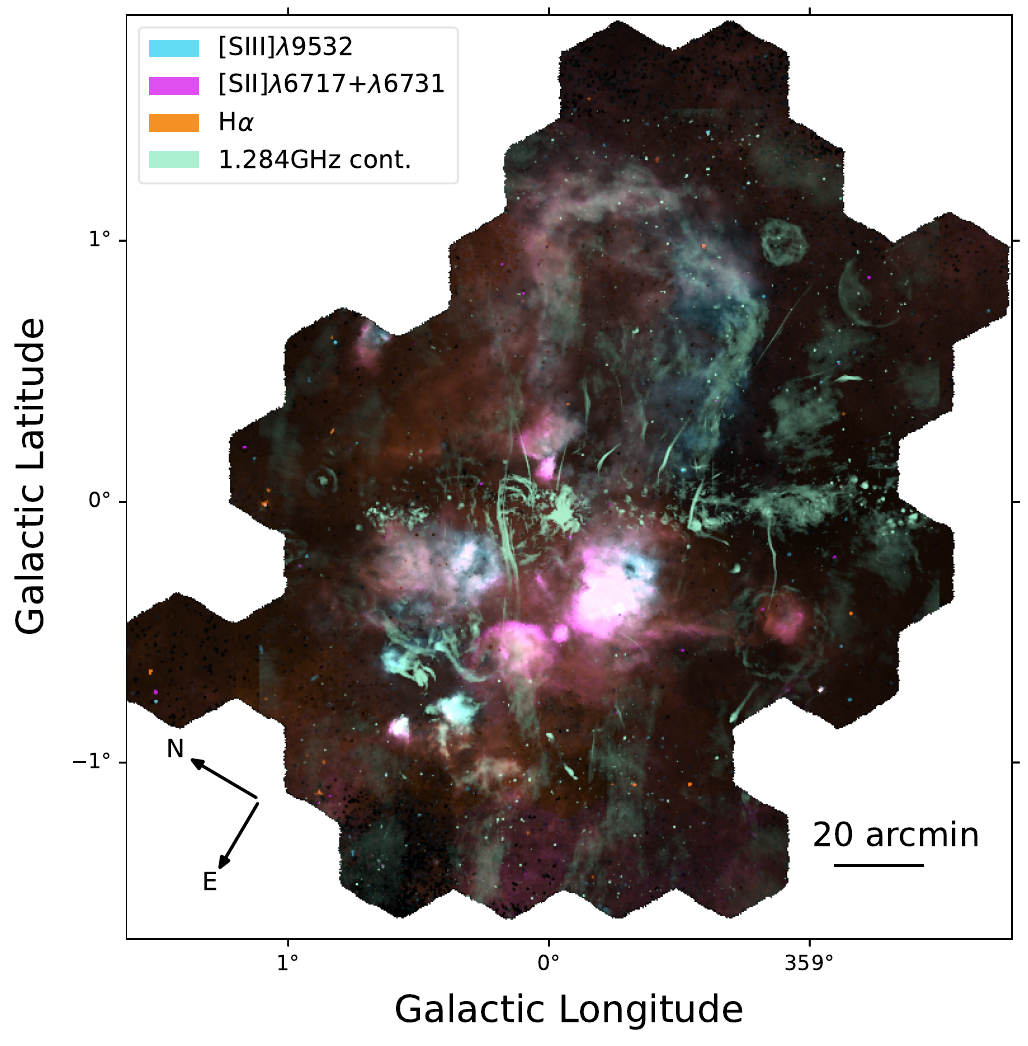}
    \caption{The Galactic Center, as seen combining optical ionized gas and radio continuum emission. 1.284 GHz radio continuum emission imaged by MeerKAT \citep[green;][]{Heywood2022} fills the plane, with narrow filaments extending vertically. LVM cannot directly probe the heavily extincted midplane, but bright line emission is apparent above and below, showing a variety of emission line ratios as traced by \ha\ (orange), \sii\ (pink) and \siii\ (blue).  Clearly this field produces projections of many objects blended along this complicated line-of-sight. For annotation and labeling of the GCL, see Figure \ref{fig:subregions}.} 
    \label{fig:overview}
\end{figure*}

\begin{figure*}
    \centering
    \includegraphics[width=0.8\linewidth]{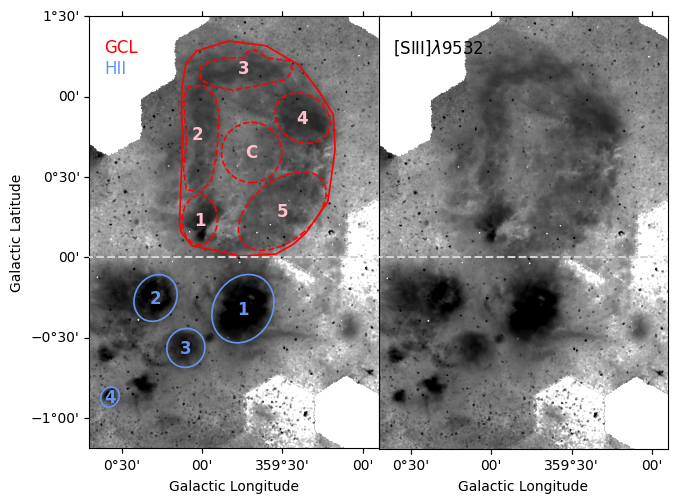}
    \caption{\siii\ suffers least from extinction, and provides the most complete and accurate view of the morphology of ionized gas in the direction of the Galactic Center. The direct view of the GCL (right) traces out clearly the full structure of the lobe, as annotated and divided into subregions (left).  The `Integrated' full lobe (solid red line) has further been subdivided into components (red dashed lines) labeled GCL-1 to GCL-5, along with a single smaller region directly in the center (GCL-C). Four bright \hii\ regions (labeled HII-1 to HII-4) are also selected at b$<$0$^{\circ}$ (blue lines).  GCL-1 and HII-1 to HII-4 have tentative designations as \hii\ regions based on nebulosity in photographic plate surveys \citep{Sh2_1959ApJS....4..257S, RCW_1976A&AS...25...25D}. The Galactic Plane is marked with a horizontal grey dashed line. }
    \label{fig:subregions}
\end{figure*}

\begin{figure*}
    \centering
    \includegraphics[width=0.9\linewidth]{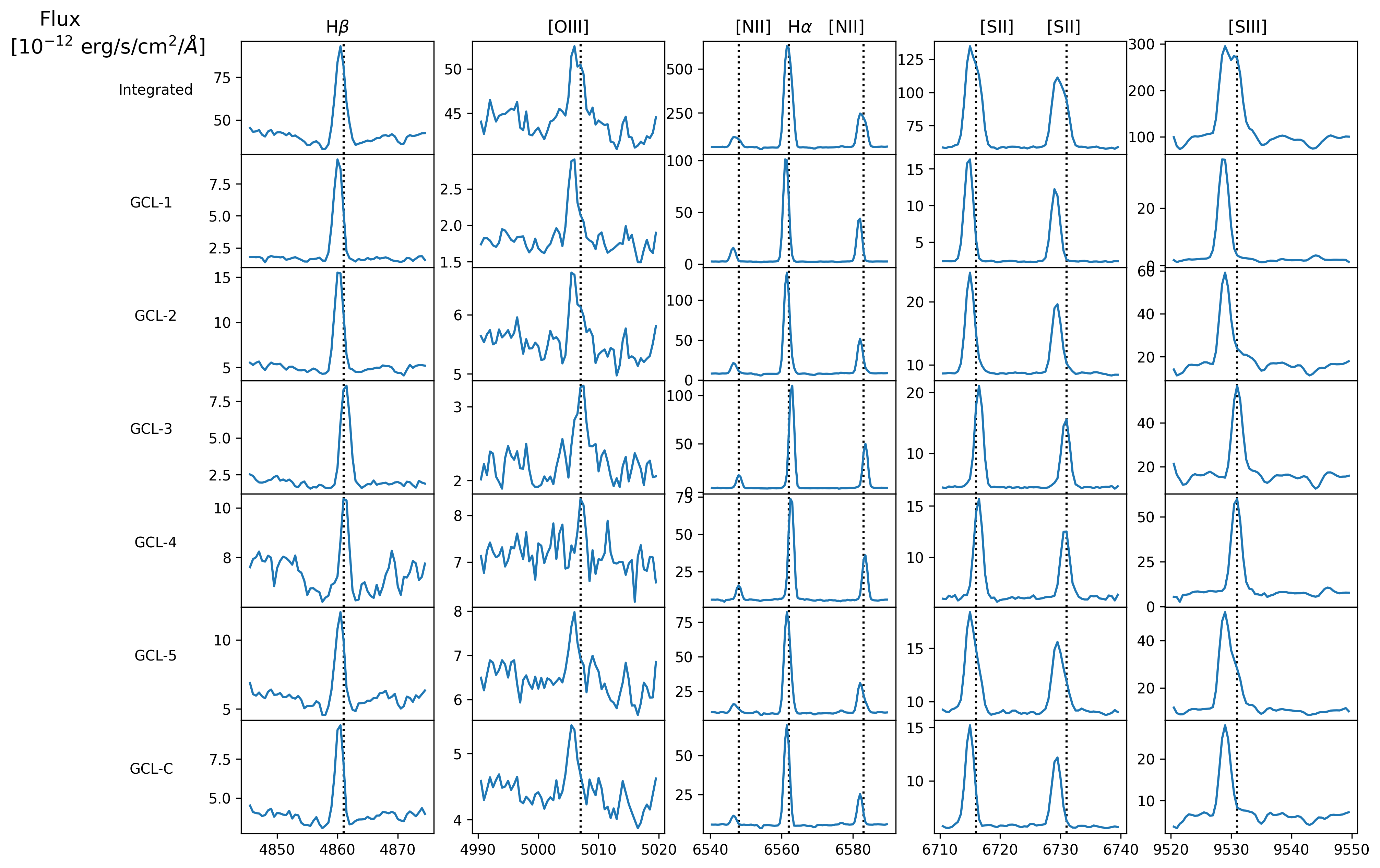}
    \caption{Line emission from the integrated spectra from each GCL subregion shown in Figure \ref{fig:subregions}. Line centers (assuming v$_{LSR}$=0~\kms) are marked with vertical dotted lines.} 
    \label{fig:spectra}
\end{figure*}

\begin{figure*}
    \centering
    \includegraphics[width=0.95\linewidth]{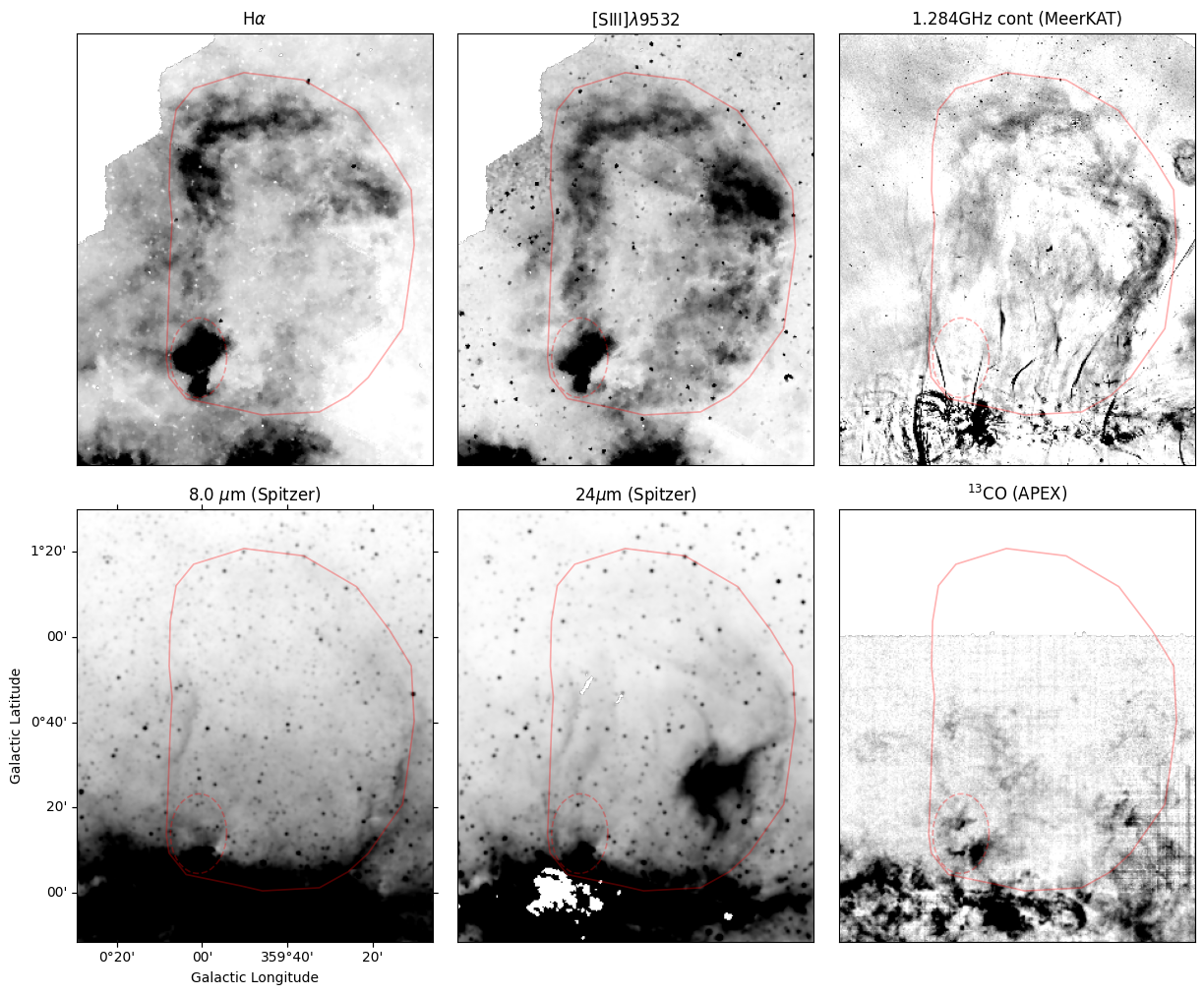}
    \caption{
    Direct multi-wavelength comparison of GCL morphology between LVM optical line emission from \ha\ and \siii\ (top left, center), MeerKAT 1.284~GHz radio continuum (top right), mid-IR dust emission in the PAH dominated 8.0~$\mu$m (bottom left) and thermal dust 24~$\mu$m (bottom center), and $^{13}$CO molecular gas (bottom right). The morphology of the \siii\ and radio lobe agrees quite well at high galactic latitudes, and is encased in an outer shell of PAH emission, with warm dust still in the region interior. CO clouds are offset, outside the GCL boundary. The GCL is marked in a solid red line, and the compact and bright GCL-1 is also marked in a red dashed line. 
    }
    \label{fig:morphology}
\end{figure*}

The GCL was discovered in 10\,GHz continuum surveys as an $\Omega$-shaped morphology, consisting of two vertical ridges connected by a diffuse arch \citep{sofue1984, Sofue1985PASJ...37..697S}.  In the mid-1980s, magnetic interpretations of Galactic Center phenomena were strongly motivated by the discovery of the Non-Thermal Filaments, such as the Radio Arc, which suggested a pervasive poloidal field component in the central few hundred parsecs \citep{Yusef-Zadeh1984Natur.310..557Y}. Within this context, the GCL was often pictured as a giant magnetic loop produced by magnetic buoyancy (\citealt{Parker1966ApJ...145..811P}), in which field lines anchored in the differentially rotating gas disk rise into the halo. 
The observed radio emission was therefore attributed to synchrotron radiation from relativistic electrons confined within the loop. 
However, synchrotron emission typically exhibits a steep spectral index ($\alpha \sim -0.7$, where $S_{\nu}\propto \nu^{\alpha}$), and multi-frequency analyses have found inconsistent results on the spectral index of the GCL ridges  \citep{Sofue1985PASJ...37..697S, Heywood2019Natur.573..235H}, while radio recombination lines confirm a thermal plasma dominates the radio emission \citep{Law2009ApJ...695.1070L, Nagoshi2019PASJ...71...80N, Anderson2024ApJ...969...43A}.

Establishing the physical conditions and Galactic location of the GCL is complicated by projection. In addition to the Galactic Center, the sight line cuts through the Sagittarius, Scutum--Crux, and Norma spiral arms. Spatial correspondences are reported between the radio continuum emission and ionized gas \citep{Law2009ApJ...695.1070L, Tsuboi2020PASJ...72L..10T}, warm dust  \citep[AFGL 5376][]{Uchida1990ApJ...351..443U, Anderson2024ApJ...969...43A}, molecular gas \citep{Uchida1994ApJ...421..505U, Veena2023A&A...674L..15V}, and X-ray emission \citep{Ponti2019, Ponti2021A&A...646A..66P}, which has been used as evidence that the GCL is associated with a multi-phase outflow \citep{Bland-Hawthorn2003}. However, in detail some of these spatial coincidences appear suspect, particularly between the X-ray and radio continuum \citep{Wang2021}, and other explanations for these alignments are also possible even if the GCL is not located at the Galactic Center.

Recent work that revisited the location and characterization of the GCL has painted an entirely different view of the GCL, with mounting evidence demonstrating that it could be a foreground object. 
Low-frequency absorption toward the GCL has been known for decades \cite[e.g.][]{Kassim1986Natur.322..522K, Reich1987}, but is only recently resolved at sufficient angular and spectral accuracy to constrain its distance.  
\cite{Nagoshi2019PASJ...71...80N} note that the GCL is seen as an absorbing structure to 74~MHz emission \citep{Brogan2003ANS...324...17B}, which along with the radio recombination lines and the thermal plasma filling the bubble favors a classification of the GCL as a giant \hii\ region.  \cite{Tsuboi2020PASJ...72L..10T} further point out that the kinematics and dust extinction towards the GCL place it at a distance of at most a few kpc, and inconsistent with the Galactic Center. The most convincing evidence was presented in joint papers that show well resolved thermal absorption \citep{Hurley-Walker2024ApJ...969...42H}, constraining the distance to $<$2~kpc, and a mid-IR comparison with other Galactic \hii\ regions \citep{Anderson2024ApJ...969...43A}, that together lead to a convincing reclassification of the GCL as an \hii\ region. 

In this paper, we present conclusive evidence to support this claim, using new observations of the GCL from the Sloan Digitial Sky Survey (SDSS-V) Local Volume Mapper (LVM). This southern sky survey includes optical integral field unit (IFU) coverage of the Milky Way Galactic Plane, ideally suited for spatially resolving ionized nebulae \citep{Kreckel2024A&A...689A.352K, Villa-Durango2025MNRAS.543.1196V, Sarbadhicary2025arXiv250708257S, Sattler2025arXiv251202802S, Hilder2025arXiv251007395H, Singh2026, Sanchez2026},  and provides new views of the emission line characteristics of ionized gas associated with the GCL. 
In Section \ref{sec:data} we present our new observations. In Section \ref{sec:results} we analyze the ionized gas morphology, the line ratio diagnostics, and the line kinematics. In Section \ref{sec:discussion} we consider the implications of these results, placing new constraints on the distance to the GCL, searching for possible stellar ionizing sources, and comparing with another local ionized gas feature (Barnard's Loop). We conclude in Section \ref{sec:conclusion}.  
Throughout this paper we use the following notation: \oiii\ to refer to \oiii$\lambda$5007, \nii\ to refer to \nii$\lambda$6583, \sii\ to refer to \sii$\lambda$6717+\sii$\lambda$6731, and \siii\ to refer to \siii$\lambda$9532.

\section{Data} \label{sec:data}

The LVM survey \citep{Drory2024} is being carried out as part of SDSS-V \citep{Kollmeier2026AJ....171...52K}, and uses a new optical IFU survey telescope built and operated at Las Campanas Observatory. It consists of four 16~cm telescopes \citep{Herbst2024}, each equipped with bundles of fibers 35.3\arcsec diameter in size, which are fed to three DESI-like spectrographs (R$\sim$4000, 3700-9800~\AA; \citealt{Konidaris2024}).  The science data is obtained from a single telescope equipped with a 1801-fiber bundle, covering a $\sim$30 arcmin diameter hexagonal field of view. Through the use of a micro-lens-array, the total fill factor of these fibers is 83\%. The three additional telescopes enable simultaneous observation of 12 spectrophotometric stars and two sky fields over the course of each 15~minute science exposure. Data reduction is carried out via a pipeline that extracts, calibrates and sky-subtracts all spectra to produce wavelength and absolute flux calibrated row-stacked spectra \citep{Drory2024}. Results presented here are produced from the tagged version v1.2.0 of the LVM data reduction pipeline\footnote{https://github.com/sdss/lvmdrp} (Meija et al. in prep). %, which is expected to form the basis of an upcoming public data release (DR), planned as DR21. 

Here we analyze observations from the Milky Way survey, which will cover $\pm18^\circ$ above and below the Galactic plane. This survey area is broken up into contiguous pointing of the LVM telescope (`tiles'), with no repeat observations or dithering. Interpolated images, such as are shown in Figure \ref{fig:overview}, are purely used for visualization, with all analysis performed directly on the observed spectra. We focus on the 43 tiles covering the Galactic Center and GCL, an area of $\sim$4.5~deg$^2$ on the sky. 

We analyze all tiles with the LVM Data Analysis Pipeline (DAP; \citealt{Sanchez2025}), which has been specifically developed to fit and subtract varying numbers of individual stars as part of 'resolved stellar population' (RSP) modeling. The main goal of this approach is to ensure the we can robustly correct for stellar continuum features and recover emission-line parameters (e.g., flux, equivalent width, systemic velocity, and velocity dispersion). We used the default input RSP model grid, which is built upon the MaStar empirical stellar library \citep{Yan2019}, consisting of approximately 30,000 spectra covering a wide range of physical parameters (see \citealt{Sanchez2025} for details). Due to the extremely high extinction towards the galactic center (A$_V > 3$~mag), we are unable to detect any faint temperature-sensitive auroral lines, and focus our study mainly on the parametric fitting of strong emission lines at less-extincted, redder wavelengths (\hb, \oiii$\lambda$5007, \ha, \nii$\lambda$6583, \sii$\lambda\lambda$6717, 6731, \siii$\lambda\lambda$9069,9532). Our approximate  5$\sigma$ \ha\ detection limit is $1\times10^{-17}$~erg~s$^{-1}$cm$^{-2}$arcsec$^{-2}$ \citep{Drory2024}, and from survey-wide comparisons with Gaia stars in the science fields we estimate $<$10\% absolute flux uncertainties. 

All Milky Way fields are observed at high shadow height ($>$1000~km) to minimize the contribution of geocoronal emission \citep{Nossal2008JGRA..11311307N}, which may contribute at a $\sim$5$\sigma$ level to our signal but does not dominate ($<$5\% of the line flux) when compared to the bright emission associated with the GCL. While this geocoronal emission should not have a strong impact on our line ratios, it may significantly bias the line centroiding of the \ha\ line. As the reddest and thus least extincted bright \siii\ line centroids appear to have small artifacts due to residual sky subtraction, our kinematic maps are obtained from the \nii\ line centers, as this bright line is detected with high S/N $>$ 20 across most of our field. All tiles are corrected for their heliocentric motions, and adjusted to the fixed Local Standard of Rest (v$_{LSR}$) as derived using the solar peculiar motion from \citealt{Schonrich2010MNRAS.403.1829S}, with typical uncertainties of $\sim$1~\kms.

The mosaic shown in Figure \ref{fig:overview} contrasts the LVM view of optical emission lines across the GC with the MeerKAT 1.284~GHz radio continuum emission \citep{Heywood2022}. The radio emission traces thermal and non-thermal sources concentrated directly along the plane ($\left|b\right| < 0.2^{\circ}$), as well as perpendicular bright and narrow filaments and shell-like supernova remnants (e.g. SNR G358.4-01.9). The LVM optical line maps are heavily extincted along the Galactic Plane itself, but at $\left|b\right| > 0.2^{\circ}$ reveals a wealth of both compact and diffuse emission. Given the high extinction, the variations in colors (and hence relative line ratios) can potentially reflect changes in both physical conditions (e.g. metallicity, ionization source) and reddening along specific lines of sight towards the emitting object (given the complicated 3D dust structures expected to be present in the spiral arms).

To simplify our analysis approach, we focus on specific sub-regions that are selected based on their emission line morphology. We use our reddest line,  \siii$\lambda$9532, to guide the identification of our sub-regions, given that it suffers the least from dust extinction and has fewer artifacts from sky subtraction and telluric corrections than the doublet line \siii$\lambda$9069. Figure \ref{fig:subregions} compares the unlabeled \siii\ map (right) with the selected subregion (left). In addition to an `Integrated' region (large, solid red line), we also consider individual components around the edge of the lobe, labeled GCL-1 to GCL-5, as well as GCL-C for the fainter center within the brighter outer ring.  Aside from the GCL itself, the only one of these subregions cataloged in the literature is GCL-1, which is identified as an \hii\ region based on nebulosity seen in photographic plate surveys, and labeled as Sh 2-17 in \cite{Sh2_1959ApJS....4..257S} and RCW 138 in \cite{RCW_1976A&AS...25...25D}, but little additional information is known.  
This object has been tentatively associated to an open cluster [DB2000] 58 \citep{DB_2000A&A...359L...9D}, which has a reported distance of 1.182 kpc (with estimated 11\% errors), E(B-V) reddening of 1.978~mag, and an age estimate of 1 Myr (MWSC 2677; \citealt{Kharchenko2013A&A...558A..53K}) derived from isochrone fitting to stellar all-sky and near-IR surveys (PPMXL, \citealt{Roeser2010AJ....139.2440R} and 2MASS \citealt{2MASS_2006AJ....131.1163S}).	Due to its treatment in the literature as a distinct \hii\ region, we consider GCL-1 as a distinct object together with the Integrated GCL in our analysis presented in Section \ref{sec:results}. 

For added context in interpreting the line ratios observed in the GCL, in Figure \ref{fig:subregions} we also isolate four \hii\ regions (in blue) at negative galactic latitudes (b $<$ 0$^\circ$), labeled HII-1 to HII-4. These have been referred to in the literature as foreground to the Galactic Center, and confirmed to be \hii\ regions by their radio recombination line emission \citep{Anderson2014ApJS..212....1A, Anderson2015ApJS..221...26A}. The coordinates in the discovery catalogs (RCW, \cite{RCW_1976A&AS...25...25D} and Sh2, \citealt{Sh2_1959ApJS....4..257S}) do not agree exactly with our objects, but there is a rough correspondence between 
HII-1 and RCW 137 (Sh2-16), 
HII-2 and RCW 141 (Sh2-20), 
HII-3 and RCW 139 (Sh2-18), 
and 
HII-4 and RCW 142 (Sh2-21). In the WISE Catalog of Galactic
HII Regions V2.3\footnote{https://astro.phys.wvu.edu/wise/}, they are listed as G359.756-00.351, G000.320-00.215, G000.120-00.556, and G000.573-00.855 (respectively). These are all seen as shadows in the low-frequency 72-103~MHz) radio data \citep{Wayth2015PASA...32...25W}, supporting their characterization as foreground \hii\ regions. 
Kinematic distance estimates place these at distances of 2-4~kpc, but with large uncertainties (see Appendix~\ref{app:blow} for more details). 

Bright, compact regions in \siii\ are also seen centered at 
(l,b)=(0$^\circ$36\arcmin,-0$^\circ$37\arcmin) and (0$^\circ$21\arcmin,-0$^\circ$48\arcmin), 
however these are not well detected in bluer lines (e.g. \sii\ or \ha, see Figure \ref{fig:overview}) likely due to heavy extinction from intervening dust clouds, and are omitted here.

Integrated spectra are extracted for each of the GCL regions identified in Figure \ref{fig:subregions}, with the key emission line features highlighted in Figure \ref{fig:spectra}. Corresponding spectra for the b $<0^{\circ}$ \hii\ regions are shown in Appendix \ref{app:blow} and Figure \ref{fig:blow_spec}. \oiii\ is the faintest line, but is detected at S/N$>$3 in all regions except GCL-4. To obtain the integrated spectra, we stacked all spectra from individual fibers residing within each region, correcting for the differences in \nii\ line-of-sight velocities between the fibers. We note that some profiles appear to have asymmetric profiles or multi-component peaks, however this is not an artifact of our stacking approach, as these features are also apparent in some individual fibers. This could it be due to background/foreground components blended along a single line-of-sight, however without higher spectral resolution it is not possible to robustly disentangle these components in the LVM data. Our integrated spectra are fit with single Gaussian profiles using the LVM DAP, in order to model and remove the underlying stellar continuum and correct for possible stellar absorption features, with the resulting line fluxes and associated propagated uncertainties analyzed in Section \ref{subsec:bpt}.

\section{Results} \label{sec:results}
Our analysis of the GCL considers the multi-wavelength morphology, characterization of the ionization source, dust extinction, emission line ratio diagnostics, and ionized-gas kinematics.

\subsection{Multi-wavelength morphology}
\label{subsec:morphology}

Literature arguments have been made for associating the GCL with the Galactic Center based on its radio continuum morphology, as it can plausibly be seen as a cap or lobe above a chimney leading downwards to the Galactic Plane. However, from the \siii\ map (Figure \ref{fig:subregions}) it is obvious that the radio continuum emission is only part of a larger, closed structure (the `Integrated' GCL, solid red contour). The cap at high galactic latitudes ($b=1^\circ10$\arcmin) can be seen to continue smoothly as a bubble structure that forms a complete loop at low galactic latitudes ($b=0^\circ15$\arcmin), which in radio continuum is obscured as it is blended in projection with emission from the Galactic Plane itself. This geometry naturally complicates interpretation of the region, and by chance the \siii\ emission appears to be bright enough to penetrate the dust screen between us and the GCL, but not bright enough that emission from the Galactic Center itself is detected. The reported southern cap of the radio continuum emission, extending from the Galactic Center towards $b=-1^\circ30$\arcmin\ and in good morphological agreement with X-ray emission \citep{Heywood2019Natur.573..235H}, shows no corresponding \siii\ emission (Figure \ref{fig:subregions}. In the context of the \siii\ emission, there is no evidence that the GCL is part of a bipolar structure centered on the Galactic Plane.

Figure \ref{fig:morphology} highlights the multi-wavelength morphology of the bubble features associated with the GCL. Comparing even \ha\ with \siii\, the completion of the lobe into a bubble was also not apparent due to the heavy extinction at low galactic latitudes providing an incomplete view of the ionized gas distribution.  In comparing these two maps, it is still clear that along specific sightlines, particularly close to GCL-1, there is still significant dust extinction seen in the \siii\ images that could hide an ionizing source for the nebula (see Section \ref{subsec:stars} for further discussion).  

The MeerKAT radio continuum 1.284~GHz emission \citep{Heywood2022} traces both thermal and non-thermal sources, and the morphology of the lobe agrees extremely well throughout the various sub-regions. There is an excess of radio continuum at the edge of the bubble (corresponding to GCL-5), and a deficit towards GCL-1 (where blending with various arcs is pronounced), however in most respects the morphological agreement is quite striking.

\cite{Anderson2024ApJ...969...43A} pointed out that also the mid-IR emission is consistent with classification of the GCL as an \hii\ region, and we highlight again here that the 8.0$\mu$m PAH emission morphology (GLIMPSE; \citealt{Benjamin2003PASP..115..953B}, \citealt{Churchwell2009PASP..121..213C}) agrees extremely well with the outer edge of the ionized bubble at both edges, and again is simply blended with the plane at the bottom. This alignment is in agreement with the expectation that PAHs are mostly ionized or destroyed close to ionizing sources but glow brightly within the photodisassociation region (PDR), as is commonly seen in Galactic \hii\ regions \citep{Allamandola1989ApJS...71..733A, Berne2022PASP..134e4301B}. 
Thermal 24$\mu$m dust emission (MIPSGAL; \citealt{Carey2009PASP..121...76C}) is also seen associated with this bubble, and overlaps with the particularly highly extincted bottom right edge of the bubble that is faint in the \ha\ imaging (GCL-5 in Figure \ref{fig:morphology}). This compact dust cloud is cataloged in the literature as AFGL 5376 \citep{Cox1989IAUS..136..121C, Uchida1990ApJ...351..443U, Uchida1994ApJ...421..505U}.

The $^{13}$CO molecular gas, observed as part of the SEDGISM survey \citep{Schuller2017A&A...601A.124S, Schuller2021MNRAS.500.3064S}, is shown as a 0th moment map that is integrated over the velocity channels from -9.5 km~s$^{-1}$ to 19.5 km~s$^{-1}$. This is a wider velocity range than was considered in \cite{Veena2023A&A...674L..15V}, where they focused on channels from -3.5 to 3.5 that traced the GCL structure, however the wider velocities also include diffuse molecular gas emission that could represent the approaching and receding sides of the GCL.  
It remains very challenging to disentangle Galactic Plane emission from what may be associated with the GCL at low galactic latitudes. 

The bright subregion GCL-1, identified in archival catalogs as an \hii\ region, also shows an off-center peak in dust emission (both 8.0$\mu$m and 24$\mu$m), and the suggestion of coincident radio continuum emission (although this region is in a forest of radio filaments and poorly highlighted with the dynamic range shown in Figure \ref{fig:morphology}). 
It is also coincident with the most prominent $^{13}$CO cloud (coincident with GCL-1 and marked in Figure \ref{fig:morphology} with a red dashed line), with a complicated molecular gas morphology that could be reflecting early stages of stellar feedback carving out bubbles and channels. 
If this is an \hii\ region in its own right, either distinct or linked to the GCL, then it has a clear blister morphology with the bottom-right edge still heavily embedded while the upper-left side at higher galactic latitudes can freely expand into the surrounding ISM. 

We note that the ionized gas morphology as traced by \ha\ and \siii\ is not entirely what is expected for an \hii\ region, as both show distinctive ring- or shell-like morphologies in the GCL. All the foreground b$<0^\circ$ \hii\ regions are more compact and centrally concentrated, showing morphologies typical of Galactic \hii\ regions. This hollow structure is particularly unusual for the higher ionization zone of a nebula, traced here by \siii\ but in the literature commonly traced by \oiii\, as these lines are generally concentrated towards the ionizing source \citep{Kewley2019}. However, the ionized gas morphology also necessarily reflects the underlying ISM morphology, so this may simply reflect pre-existing structures present in the ISM in this region. This idea will be discussed in more detail in Section \ref{subsec:barnardsloop}.

\subsection{Characterizing ionization conditions} 
\label{subsec:bpt}

In extragalactic studies of ionized nebulae, the determination of whether or not an object is photoionized typically comes from the consideration of multiple emission line ratio diagnostics. Historically, \cite{Baldwin1981} showed that \oiii/\hb\ compared to \nii/\ha\ provides a robust way to separate photoionized \hii\ regions from planetary nebulae and shocked supernova remnants. Later work included additional diagnostics, primarily \sii/\ha, to aid in the characterizaiton of the ionizing source. These so-called BPT diagrams are typically divided by demarcation lines determined by theoretical photoionization modeling of integrated \hii\ regions \citep{Kewley2001ApJ...556..121K}, or empirical characterization of integrated galaxies \citep{Kauffmann2003}. 
The placement of an ionized nebula below these diagnostic demarcations does not guarantee the object is photoionized \citep{Sanchez2020ARA&A..58...99S, Sanchez2025A&A...704A.145S}, but it commonly used to interpret extralactic observations. 
These diagnostics are challenging to apply to Galactic sources, as they require an integrated view of the entire nebula to accurately recover emission lines from both the inner and outer ionization zones of the nebula \citep{Kreckel2024A&A...689A.352K}. With its wide-area IFU coverage, LVM is one of the first instruments that can systematically perform this integrated analysis on Galatic \hii\ regions.

\begin{figure}
    \centering
    \includegraphics[width=0.99\linewidth]{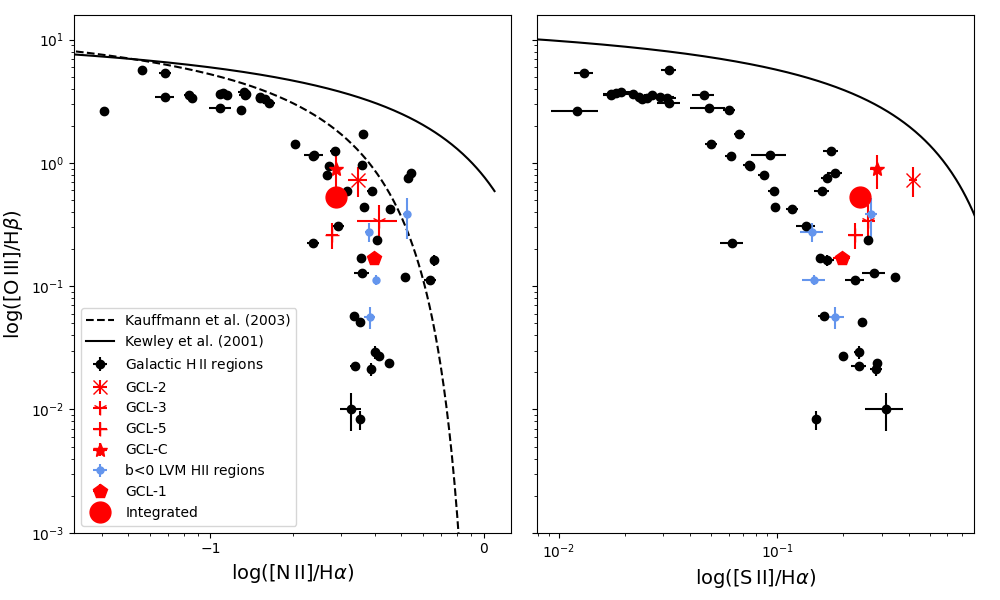}
    \caption{BPT diagnostic diagrams compare the measured \oiii/\hb\ with \nii/\ha\ and \sii/\ha\ for the Integrated GCL and each integrated subregion (red), along with b$<0^\circ$ \hii\ regions (blue), as labeled in Figure \ref{fig:subregions}. These are compared to a set of Galactic \hii\ regions observed by the DESIRED sample \citep{jemd_desired}. All regions are consistent with photoionization when compared to the \cite{Kewley2001ApJ...556..121K} demarcation line, and most are also consistent with the empirical \cite{Kauffmann2003} line. }
    \label{fig:BPT}
\end{figure}

In Figure \ref{fig:BPT} we show these two BPT diagrams for the GCL, including all subregions, and the set of four foreground b$<$0$^\circ$ \hii\ regions in Figure \ref{fig:subregions} as representative of typical Galactic \hii\ regions. We require S/N$>$3 for all lines, resulting in the exclusion of GCL-4, where \oiii\ is not significantly detected. Note that because all lines are very close in wavelength, no extinction corrections are applied when calculating these line ratios. Even a very high extinction of A$_V$ = 4~mag would result in changes to these line ratios well within the error bars. 
For some context and comparison we overlay data from a set of Galactic \hii\ regions in the DESIRED sample \citep{jemd_desired}, but note that these may not recover flux from the entire regions. All regions are consistent with photoionization when compared to the \cite{Kewley2001ApJ...556..121K} demarcation line, and most are also consistent with the empirical \cite{Kauffmann2003} line. In particular, the Integrated and GCL-1 regions are both fully consistent with  photoionization, and in good agreement with the DESIRED Galactic \hii\ regions and with the four b$<0^\circ$ \hii\ regions.  From this analysis, any extragalactic study would conclude that the GCL is an \hii\ region.

\subsection{Dust reddening and extinction} \label{sec:dust}

Due to the high amount of dust extinction evident in this field, comparison of lines with wider wavelength separations and determination of absolute line luminosities is very sensitive to the exact reddening correction applied. Typically this is performed using the \ha/\hb\ Balmer decrement, assuming case B recombination, to constrain the normalization of a given extinction or attenuation law, however these often contain an additional scaling term parameterized by R$_V$. This reflects the steepness of the chosen curve, and the value has been shown to vary significantly from diffuse to dense ISM conditions \citep{Zhang2025Sci...387.1209Z}. With the addition of a third hydrogen recombination line, it is possible to also constrain this R$_V$ parameter, and ensure even more robust recovery of reddening corrected parameters. 

\begin{figure}
    \centering
    \includegraphics[width=0.99\linewidth]{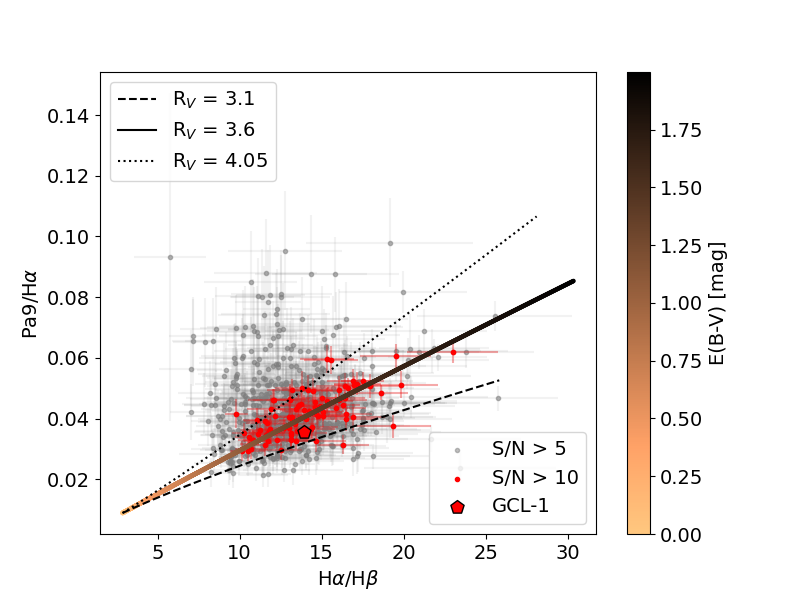}
    \caption{Comparison of Pa9/\ha/ with \ha/\hb\ to place constraints on the shape of the dust extinction law. Individual fibers with modest (S/N$>$5; grey) and high (S/N$>$10; red) confidence detections of Pa9, along with the integrated measurement from GCL-1, are compared with predictions from the \cite{Fitzpatrick1999} extinction curve with a range of R$_V$ values (dotted, dashed, and solid lines). The solid line is further colored to show how the lines ratios would vary with changes in E(B-V).  We find that R$_V$=3.6 gives the best agreement, and corresponds to values of E(B-V) = 1.0-1.5.
    }
    \label{fig:balmer_paschen}
\end{figure}

With LVM, at the blue end we cover many additional Balmer lines, however they are all not detected due to the high dust extinction. Luckily, from the wide wavelength coverage to the red we observe the Paschen 9 recombination line at $\lambda$9229 (Pa\,9). Due to the faintness of this line, we find it is primarily detected within a few individual fibers that contain the brightest regions of the GCL, and within the high surface brightness GCL-1 region. In Figure \ref{fig:balmer_paschen} we compare Pa9/\ha\ to \ha/\hb\ for all individual fibers with S/N$>$5 (grey) and S/N$>$10 (red). As most extinction laws are very similar in shape in the optical, we adopt the \cite{Fitzpatrick1999} extinction curve and overplot lines for fiducial values of R$_V$=3.1 (dashed, typical of diffuse conditions; \citealt{CCM}) and R$_V$=4.05 (dotted, typical of denser conditions; \citealt{Calzetti2000}). We find our high confidence detections and the GCL-1 region fall between these two regimes, consistent with R$_V$=3.6, and adopt this value for the remainder of our analysis. This is typical for translucent regions in the ISM, and may be associated with changes in the dust size distribution \citep{Zhang2025ApJ...979L..17Z}. In addition, we note that this is significantly different from the value of R$_V$=2 that has previously been determined for the Galactic Center \citep{Popowski2000ApJ...528L...9P, Nishiyama2008ApJ...680.1174N}, providing additional evidence for more local, star-forming ISM conditions in the GCL.

Adopting this fixed value of R$_V$=3.6, we use $pyneb$ \citep{Luridiana2012} to convert our \ha/\hb\ line ratios to a measurement of the reddening, E(B-V), in each integrated region. We find values ranging between 0.5 to 1.5~mag, with most at $\sim$1.2~mag (A$_V$ = 4.3~mag), in very good agreement with estimates from broadband colors in the direction of the GCL \citep{Tsuboi2020PASJ...72L..10T}. 

\subsection{Oxygen abundance diagnostics}
%Emission line ratio diagnostics} 
\label{subsec:abundances}

One of the confusing aspects of the GCL is its low reported value of \te$\sim$3500~K based on radio recombination line measurements \citep{Law2009ApJ...695.1070L, Anderson2024ApJ...969...43A}. Unfortunately, we are unable to detect any of the faint auroral lines (e.g. \nii$\lambda$5755, \sii$\lambda$6312, \oii$\lambda\lambda$7320,7330) that would allow us to directly measure \te. From the \sii$\lambda$6717/\sii$\lambda$6731 density diagnostic we are also unable to provide strong constraints on the electron density, n$_e$, with all measurements consistent with the low-density limit (n$_e < 100$cm$^{-3}$) within the uncertainties. Instead, we consider our strong-line ratios that are sensitive to changes in \te\ and metallicity \citep{Kewley2019}, within the context of the DESIRED sample of Galactic \hii\ regions. 

\begin{figure}
    \centering
    \includegraphics[width=0.95\linewidth]{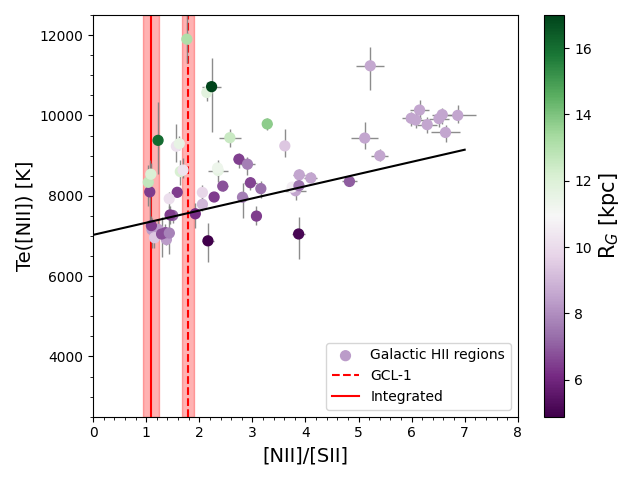}
    \includegraphics[width=0.95\linewidth]{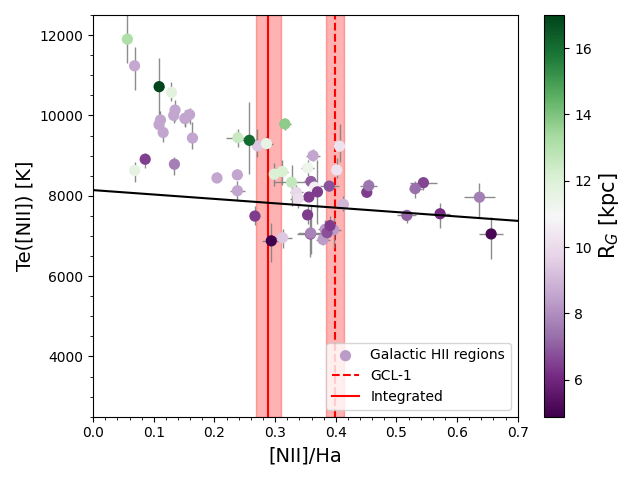}
    \caption{\te\nii\ as a function of \nii/\sii\ (top) and \nii/\sii\ (bottom) for a sample of Galactic \hii\ regions \citep{jemd_desired}. Points are colored by their Galactocentric radius (R$_G$), and to guide the eye a linear fit (black) is performed for regions within the solar radius (R$_G < 8.5$~kpc).  Our Integrated GCL (solid red line) and GCL-1 (dashed red line) measurements are in agreement with the Galactic \hii\ regions, and consistent with a Te $\sim$7000-8000~K. }
    \label{fig:Te_inferred}
\end{figure}

Figure \ref{fig:Te_inferred} shows \te(\nii) as a function of \nii/\sii\ and \nii/\ha\ for a sample of $\sim$40 Galactic \hii\ regions, colored by their Galactocentric radius (R$_G$). Here, the distances are taken from \cite{MendezDelgado:2022a}, and based on the identification of ionizing stellar sources from Gaia that have robust parallax measurements. 
Each line ratio shows a correlation with \te, which we highlight by carrying out a linear fit to all objects with R$_G < 8.5$~kpc (interior to the solar radius). We mark the values for the Integrated GCL (solid red line) and GCL-1 (dashed red line), which are slightly offset from each other but would both overlap with the sample of Galactic \hii\ regions. These line ratios correspond to \te$\sim$7000-8000~K, although we emphasize that there is also significant scatter ($\sim$1000~K) in the Galactic \hii\ regions at fixed line ratios, making this a weak diagnostic of \te. \te$\sim$7000-8000~K is slightly lower than the value typically associated with solar radius from the Orion Nebula (8300~K; \citealt{MendezDelgado:2022a}), but significantly higher than the value reported from radio recombination lines. 

As \te\ is typically of interest as a diagnostic of the gas-phase metallicity, 12+log(O/H), with lower \te\ corresponding to higher 12+log(O/H) due to more efficient metal cooling, we also consider the metallicity of each region using integrated strongline abundances. 
From our constraints on the dust reddening (Section \ref{sec:dust}), we correct all detected lines for extinction and calculate the metallicity using prescriptions that focus on only the red part of the spectrum. We are able to apply the \cite{Pilyugin2016} S-calibration and the N2S2 \cite{Dopita2016} calibration  to both the Integrated GCL and GCL-1. These two calibrations are generally well correlated \citep{Groves2023}, however the  N2S2 prescription is typically 0.2~dex higher than the S-calibration, partly because of the different zero-points (S-calibration and N2S2 adopt different solar values of $\rm 12+\log(O/H)=8.69$ and 8.77, from \citealt{Asplund2021A&A...653A.141A} and \citealt{Nieva2012}, respectively). For the Ingegrated GCL (GCL-1) we find a value of 12+log(O/H) from N2S2 of  8.67$\pm$0.03  (8.93$\pm$0.01), and from the S-calibration we find a value of 8.50$\pm$0.03 (8.63$\pm$0.01). We discuss the implications of our \te\ and metallicity measurements on the distance to the GCL in Section \ref{subsec:dist}.

\subsection{Kinematics} \label{subsec:kinematics}

\begin{figure}
    \centering
    \includegraphics[width=0.99\linewidth]{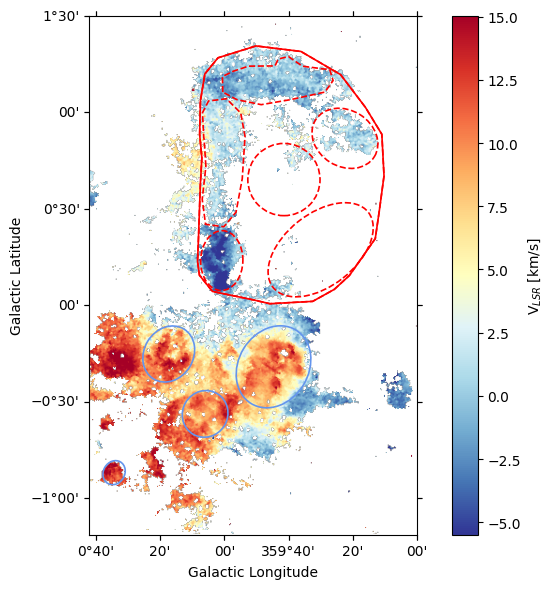}
    \caption{A map of the ionized gas velocity based on the \nii\ line centroid, which is masked at S/N$<$20. The GCL (red line) shows a uniform velocity structure, which is offset from the b$<0^\circ$ \hii\ regions (blue lines). }
    \label{fig:NII_vel}
\end{figure}

Figure \ref{fig:NII_vel} shows the ionized gas kinematics across our field, as measured from the \nii\ line, and highlights the narrow range of velocities (v$_{LSR}$ $\sim$ -5 to 5 \kms) across the GCL. We see no obvious signs of expansion, or gradients in Galactic Longitude (which might be associated with Galactic rotation at the Galactic Center), in good agreement with what has been previously reported from the RRL kinematics \citep{Nagoshi2019PASJ...71...80N}. These kinematics are also in rough agreement with what is reported from the $^{13}$CO line \citep{Veena2023A&A...674L..15V}. 
The measured intrinsic velocity dispersion measured in the shell, as constrained by \nii\ and corrected for the instrumental line spread function, is typical for \hii\ regions (10-15 \kms). We see no prominent indications of further line broadening that could be measured at our S/N and spectral resolution.

We also note that the complex of bright line emission at b$<0^\circ$, including the four \hii\ regions isolated in Figure \ref{fig:subregions}, span a very narrow range of velocities and is offset to higher velocities (v$_{LSR}$ $\sim$ 10-15 \kms). This suggests that this complex is at a slightly different distance than the GCL, but also not at the Galactic Center (as there are no Galactic rotation signatures).

\section{Discussion} \label{sec:discussion}

Taking into account our new constraints on the optical emission line characteristics of the GCL, we reconsider what, and where, this object is.

\subsection{How far away is the GCL?} \label{subsec:dist}

While the morphology and geometry of the GCL has long led to the assumption that it is associated with an outflow located at a distance of $\sim$8.5~kpc in the Galactic Center \citep{Bland-Hawthorn2003}, recent work has convincingly argued that it is instead a foreground \hii\ region \citep{Nagoshi2019PASJ...71...80N, Tsuboi2020PASJ...72L..10T, Anderson2024ApJ...969...43A, Hurley-Walker2024ApJ...969...42H}. Based on the optical emission line diagnostics we confirm that the GCL is photoionized, and consistent with other Galactic \hii\ regions. 

\begin{figure*}
    \centering
    \includegraphics[width=0.99\linewidth]{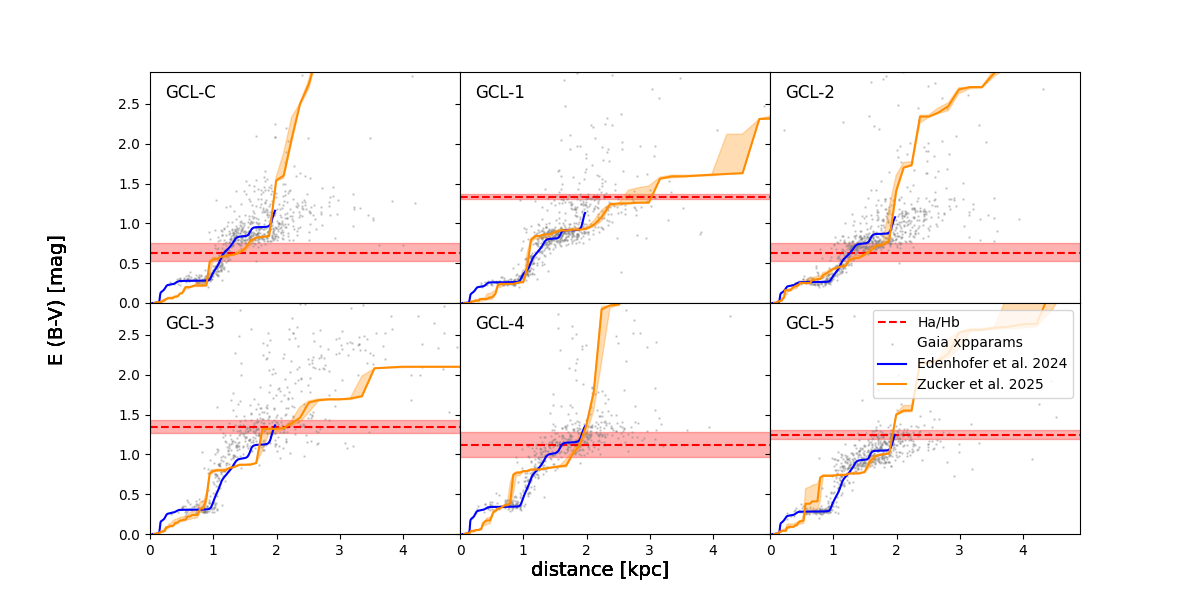}
    \caption{E(B-V) reddening as a function of distance towards each of our regions (see Figure \ref{fig:subregions}), comparing measurements from stars and from gas. Our measurements from the Balmer decrement (red) integrated across each sub-region are compared with 3D dust maps derived from stars (orange, \citealt{Zucker2025ApJ...992...39Z}; blue, \citealt{Edenhofer2024A&A...685A..82E}), and constrain the distance of the GCL to $\sim$2~kpc. The widths of each line indicate the 1$\sigma$ uncertainties. We also show reddening along sightlines to individual Gaia sources (grey points).  }
    \label{fig:gaia_dust}
\end{figure*}

The development of new 3D dust mapping of the Milky Way has provided a remarkable new way to understand the ISM structures in absolute distance units, without the uncertainties associated with degenerate kinematic constraints. Particularly along the line of sight directly towards the Galactic Center, this offers a completely new opportunity to constrain the physical location of the GCL. In Figure \ref{fig:gaia_dust} we compare our reddening measurements (E(B-V); in red) from the \ha/\hb\ Balmer decrement in each integrated region with the integrated reddening as a function of distance along that specific line of sight, as measured in two of the recent 3D dust maps and accessed with the python package $dustmaps$ \citep{Green2018JOSS....3..695M}. The DECaPS map from \cite{Zucker2025ApJ...992...39Z} provides the most suitable comparison (orange), as it focuses on the Southern Galactic Plane using a combination of optical and infrared stellar catalogs to constrain the dust distribution out to maximum distances of 10~kpc. We also include constraints from the \cite{Edenhofer2024A&A...685A..82E} dust map (blue), which is determined using the Gaia BP/RP spectra, but is only available out to distances of 2~kpc. For context, we include measurements towards individual Gaia stars (grey points), as determined from the Gaia BP/RP spectra \citep{xpparams_2023MNRAS.524.1855Z}, that are within 9\arcmin\ of the region center.  

Strikingly, the reddening measured from the ionized gas along all sightlines is consistent with a distance of $\sim$2~kpc, particularly when compared to the DECaPS maps. In all tracers, it is clear that there is a sheet of dust at $\sim$1~kpc, causing a jump in the reddening, but what emerges in the DECaPS maps is a second sheet at $\sim$2~kpc that causes any more distant emission to suffer extremely high E(B-V)$>$2 reddening. This is particularly pronounced along certain sightlines (e.g. GCL-C, and GCL-2, -4 and -5). GCL-1 appears to exhibit slightly higher E(B-V) than would be predicted from the dust maps, however from the multi-wavelength imaging (Figure \ref{fig:morphology}) it is clear that there is a localized dust concentration associated with this region that could be expected to produce locally elevated E(B-V). From these dust constraints, it is clear that the GCL cannot be at the Galactic Center, and must be at a distance of at most 2~kpc from the sun. 

\begin{figure}
    \centering
    \includegraphics[width=0.99\linewidth]{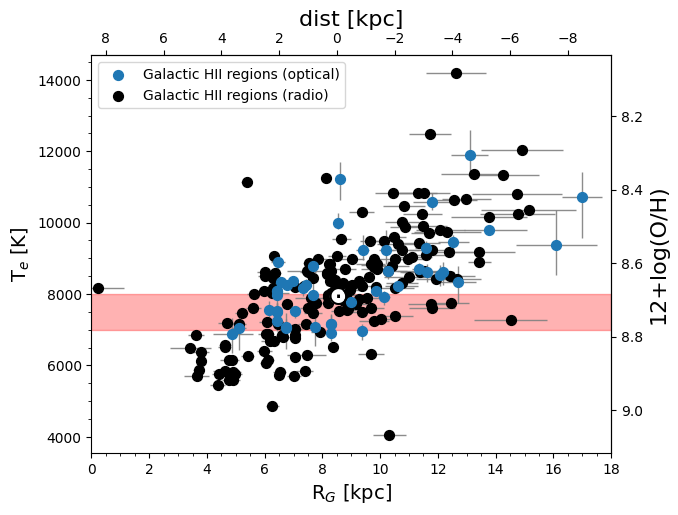}
    \caption{\te\ as a function of Galactocentric radius (R$_G$), for a sample of optical \nii\ \citep[blue, ][]{jemd_desired} and radio recombination \citep[black,][]{Wenger2019} line observations of \hii\ regions. The alternate y-axis converts these \te\ measurements to a metallicity constraint by adopting the prescription that assumes the presence of temperature fluctuations \citep{Martinez-Hernandez2026}. Our constraint for the GCL (red band) places it at a distance of less than 4~kpc. 
    }
    \label{fig:MW_gradient}
\end{figure}

While our measurements of \te\ and 12+log(O/H) are not as precise as E(B-V), these can also provide rough constraints on the radial location of the GCL given the established temperature and metallicity gradients in the Milky Way. In Figure \ref{fig:MW_gradient} we show the distribution of the optically determined \te\nii\ as a function of distance for the DESIRED \hii\ regions (also shown in Figure \ref{fig:Te_inferred}), along with the \te\ determined from radio recombination line observations \citep{Wenger2019}. The alternate y-axis shows a conversion of \te\ to 12+log(O/H) using the relation determined by \cite{Martinez-Hernandez2026}, assuming the presence of temperature fluctuations (t$^2 > 0$). 
Our  estimation of \te$\sim$7000-8000~K, based on the emission line ratios presented in Section \ref{subsec:abundances}, is consistent with R$_G$=4-10~kpc (distances of 0-4~kpc), although we note that this \te\ is physically inconsistent with the observed narrow radio recombination line profiles \citep{Law2009ApJ...695.1070L, Anderson2024ApJ...969...43A}. However, \te$\sim$7000-8000~K is also in very rough agreement with the strongline metallicities (ranging from 8.5-8.8) that are inferred, with large uncertainties introduced depending on the metallicity prescription and region considered. While there are limited measurements available at distances beyond 4~kpc, again our constraints strongly rule out a location directly at the Galactic Center. 

\subsection{What ionizing stars might be associated with the GCL?} \label{subsec:stars}

One piece of the puzzle still missing is the robust identification of an ionizing source for this nebula. \cite{Anderson2024ApJ...969...43A} suggested various candidates, particularly in regions close to the 24$\mu$m bright dust feature, but none of them have a firm classification as an O type star. Based on our extinction corrected \ha\ flux, and assuming a distance of 2~kpc (see Section \ref{subsec:dist}), we recover a total \ha\ luminosity (accounting for the 83\% filling factor) for the GCL Integrated region of 1.5$\times$10$^{37}$ erg~s$^{-1}$. Given its $\sim$35~pc diameter size, this suggest that the GCL also approximately follows the \hii\ region size-luminosity relation, as measured in extragalactic systems \citep{Barnes2026A&A...706A..95B}. If we were to consider GCL-1 alone as an \hii\ region, as it is reported in the literature photometric nebular catalogs as RCW 138, then it has a total \ha\ luminosity of $3.9\times10^{36}$ erg~s$^{-1}$.   These luminosities are consistent with the ionizing photon flux expected from an individual O 5.5 V star (log Q$_{0}$ = 49.1; \citealt{Martins2005A&A...436.1049M}), or at least 10 B type stars \citep{Sternberg2003ApJ...599.1333S}.

\begin{figure*}
    \centering
    \includegraphics[width=0.9\linewidth]{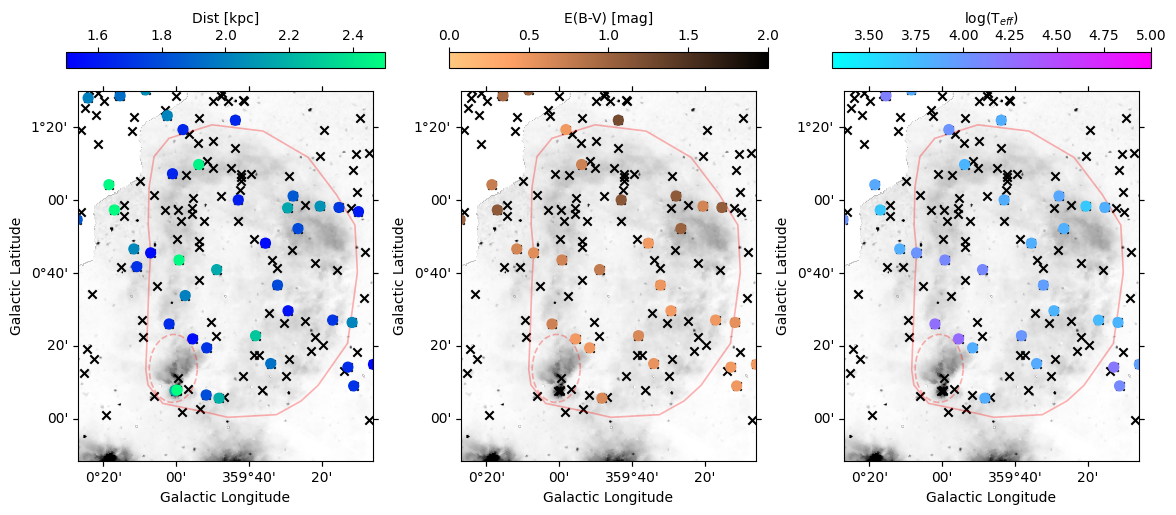}
    \caption{The locations of OB stars from \cite{Zari2021A&A...650A.112Z} compared to the morphology of the GCL (greyscale, traced here by \siii). Stars with parallax distances between 1.5 and 2.5~kpc are colored by their distance (left), reddening (E(B-V), center) and effective temperature (T$_{eff}$, right). All stars with distances between 0-3~kpc are also shown in black (crosses). Interestingly, most stars avoid the center of the GCL, and none are characterized as O type (log T$_{eff}$ $>$ 4.5). }
    \label{fig:stars2kpc}
\end{figure*}

Most literature catalogs of massive stars do not cover this part of the sky \citep{ Clark2021A&A...649A..43C, Pantaleoni2025MNRAS.543...63P}. 
In a renewed effort to search for O and B type stars that could be ionizing the GCL, we consider a recent catalog of OB stars \citep{Zari2021A&A...650A.112Z} and select all objects where the distance (as inferred from the parallax) is between 1.5 and 2.5~kpc. A subset of these (in color) have additional characterization of the stellar parameters ($\log\,g$, log T$_{eff}$) available based on SDSS-V Milky Way Mapper (MWM) spectroscopy (BOSSNet; \citealt{Sizemore2024AJ....167..173S}). Figure \ref{fig:stars2kpc} shows their location with respect to the GCL, colored by their distance (right) and when available also their reddening (E(B-V), center) and effective temperature (T$_{eff}$, right). None of them are characterized as O type stars (which would require log T$_{eff} > 4.5$), and there is no apparent concentration or cluster in the center of the lobe. A single star, Gaia DR3 4060180404854339328, is located slightly more central to the GCL, with E(B-V)$\sim$0.8~mag, consistent with what is obtained from the Balmer decrement, however with only T$_{eff}$$\sim$12,000~K it would seem like a B type star that is insufficient to power this entire nebula. 
Interestingly, there is a suggestion that these OB stars are instead clustered around the outer ring of the GCL. This is even more apparent when including all OB stars with distances between 0 and 3~kpc (black crosses), as they strongly avoid the location directly at the center of the GCL. 

An interesting potential candidate for the ionization source of the lobe is the cluster associated with GCL-1, [DB2000] 58. In \cite{Kharchenko2013A&A...558A..53K}, isochrone fitting of this young open cluster resulted in an age of $\sim$1~Myr, and there are clearly B stars identified by \cite{Zari2021A&A...650A.112Z} associated with it. While the  distance estimate of 1~kpc is slightly closer than what we infer, the cluster is clearly heavily embedded, with E(B-V)= 1.978~mag, which is one of the highest reddenings reported in that work and may result in a distance bias. However, the location of this cluster and of GCL-1 is clearly off-center from the GCL and located instead on the edge, making the interpretation of the geometry challenging. 

\subsection{A possible counterpart to Barnard's Loop?}
\label{subsec:barnardsloop}
\begin{figure*}
    \centering
    \includegraphics[width=0.95\linewidth]{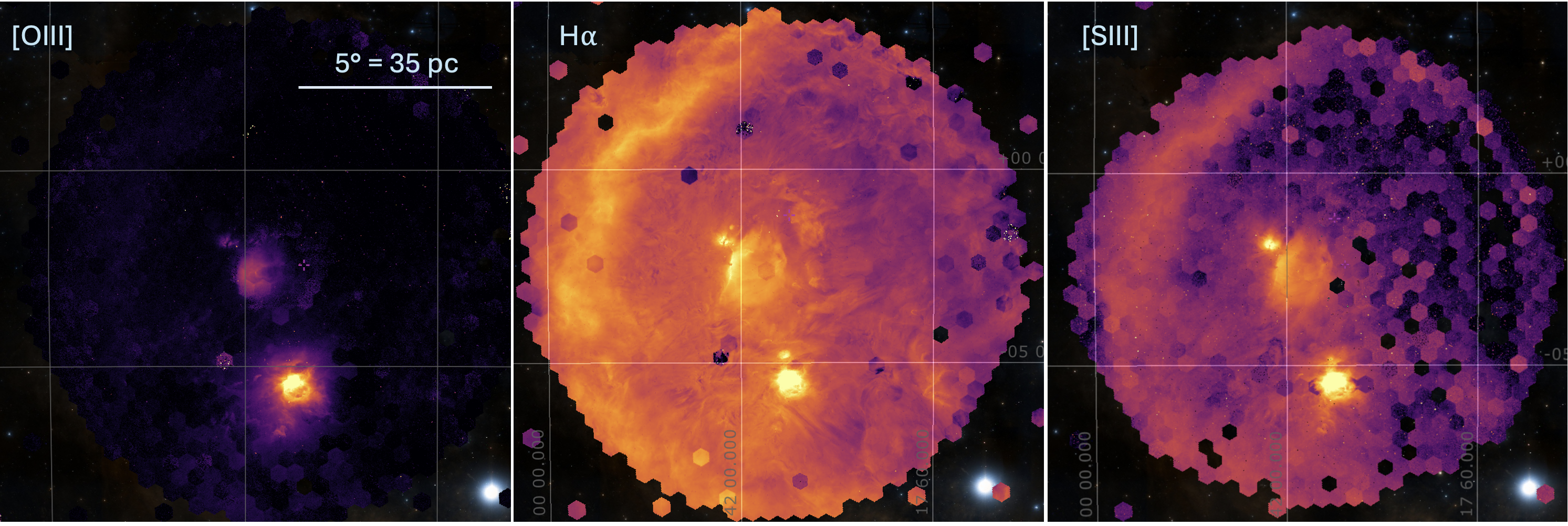}
    \caption{Barnard's Loop in LVM: \oiii\ (left), \ha\ (center), \siii\ (right). Covering nearly $15^\circ \times 15^\circ$ across the sky ($\sim$100~pc in diameter), this ionized ring is associated with the O type stars powering M42, NGC~2024 and IC434 \citep{Kreckel2024A&A...689A.352K}. }
    \label{fig:BL}
\end{figure*}

The unusual \siii\ morphology, which is not concentrated on a ionizing source but instead forms a ring or shell, is reminiscent of what is seen in the nearby structure surrounding the Orion Nebula on large scales, known as Barnard's Loop \citep{Barnard1894PA......2..151B}. Extending over 10$^\circ$ in diameter in the sky, Figure \ref{fig:BL} demonstrates that the \siii\ morphology observed by the LVM of Barnard's Loop is strikingly similar to the GCL. While bright in \ha\ and \siii\, it is faint in \oiii, and is thought to be mainly photoionized by the massive stars located in the Orion Nebula (M42), although it does not sit directly in the center of the loop. Recent work has shown it is actually a closed bubble \citep{Ochsendorf2015ApJ...808..111O}, possibly associated with a recent ($<$1~Myr) supernova event. In a 3D kinematic study of the dust and stars associated with the region, \cite{Foley2023ApJ...947...66F} showed that Barnard's Loop corresponds to a large dust cavity, and again assert that it has been carved out by past supernovae events. Interestingly, like the GCL \citep{Law2009ApJ...695.1070L, Anderson2024ApJ...969...43A} it also has a low reported value of \te$\sim$5200~K from radio recombination line measurements  \citep{Gaylard1984MNRAS.211..149G} and \te$\sim$6000~K from a combination of \ha\ and radio continuum \citep{Heiles2000ApJ...536..335H}, which is $\sim$2300~K lower than what is reported for the Orion Nebula itself \citep{MendezDelgado:2022a}.

An examination of optical line ratio diagnostics in the LVM coverage of Barnard's Loop, compared to the \hii\ region at the young embedded Flame Nebula \citep{Kreckel2024A&A...689A.352K}, is similar to what is seen in a comparison between the GCL and GCL-1.  Using a single LVM tile (tileid=1029104) that is centered on a part of Barnard's Loop at (ra,dec) = (86.80012~deg, 1.28296~deg), we find that the ionized gas ring exhibits very low reddening (E(B-V)=0.02~mag) compared to the dust-enshrouded Flame Nebula (E(B-V)=0.54~mag). Both the GCL and Barnard's Loop have low extinction corrected \siii/\sii\ line ratios of $\sim$0.2 to 0.3, significantly lower than the ratio of $\sim$0.9 that is measured in the Flame Nebula.  As \siii/\sii\ is a proxy for changes in ionization parameter, reflecting the number of ionizing photons impacting the gas, this is also suggestive that the GCL be removed by some significant distance from the primary ionization source, as in Barnard's Loop. \sii/\ha\ is also similar in both structures, further confirming the similarities in the ionization state of the gas in both shells.  

\begin{figure}
    \centering
    \includegraphics[width=0.95\linewidth]{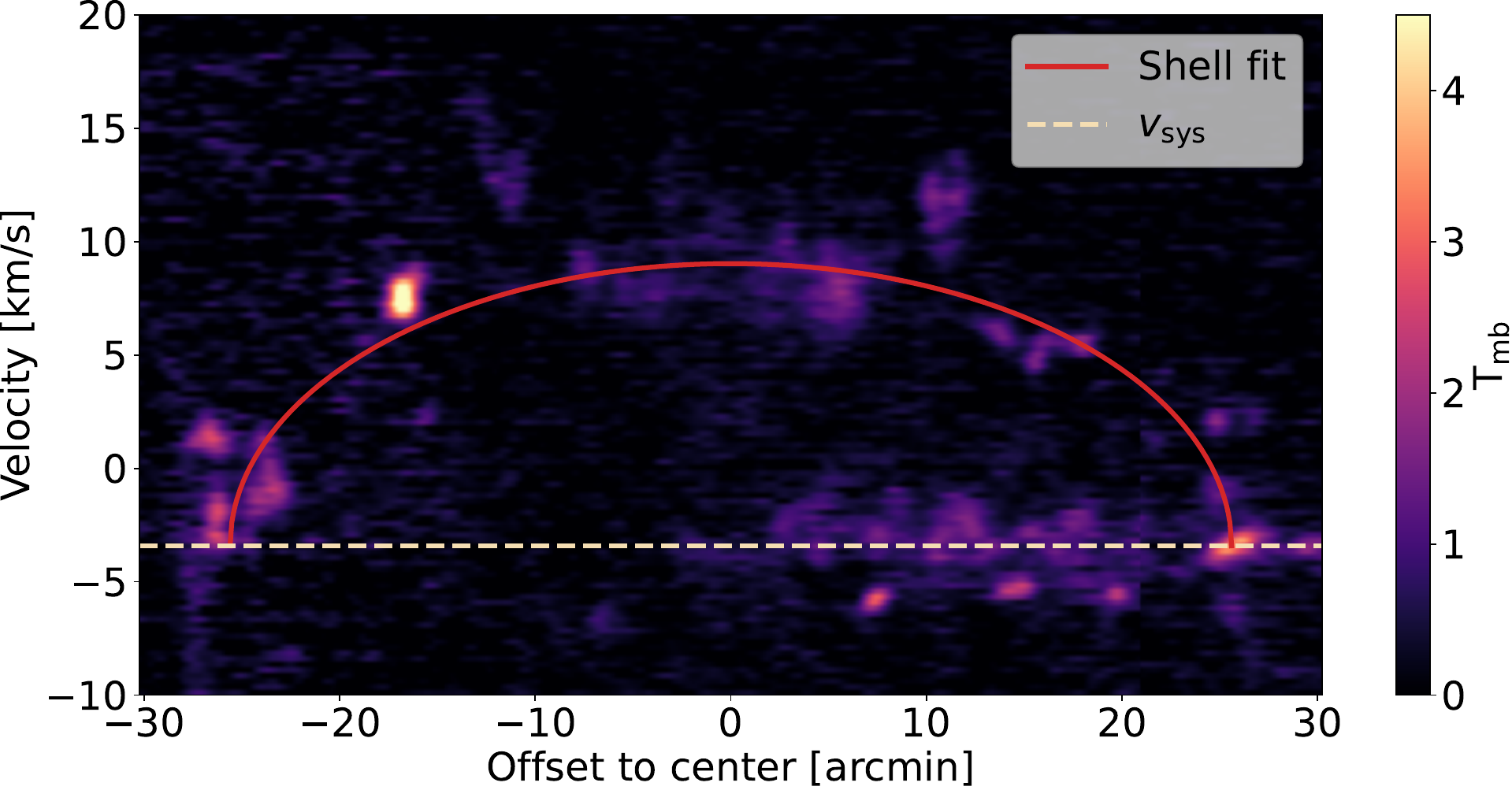}
    \caption{A position-velocity slice across the GCL center at $b = 0^{\circ}24$\arcmin\ in the $^{13}$CO SEDGISM data, as modeled by a simple expanding blister shell (Equation \ref{eqn:blister}).  }
    \label{fig:co_blister}
\end{figure}

If the GCL is at a distance of 2~kpc, then its angular size corresponds to a physical diameter of $\sim$35~pc. This is about half the size of Barnard's Loop. While we do not directly constrain an expansion velocity for the GCL due to the limited spectral resolution of the LVM, we do measure a  $\sim$10~\kms\ velocity gradient from top to bottom. This is similar to tentative broadening seen in the GCL center in radio recombination line observations from \cite{Law2009ApJ...695.1070L}, in both the center of their b=0$\fdg$45 strip and the edge of their GCL4 strip. Intriguingly, this signature is also seen in the $^{13}$CO SEDGISM data (Figure \ref{fig:co_blister}), and we model the position-velocity diagram along $b=0^{\circ}24$\arcmin\ as a blister as
\begin{equation}
\label{eqn:blister}
    v(x) = v_{\rm sys} + v_{\rm exp} \, \sqrt{1 - \left(\frac{\Delta \theta}{\theta_R}\right)^2}.
\end{equation}
With $\theta_R$ the angular radius of the bubble and  and $\Delta \theta$ the angular separation from the center $(\ell_0 = 359^\circ 39' 0'' ,\,  b_0 = 0^\circ 24' 0'')$.
The fit yields an expansion velocity of $v_{\rm exp} = 12.5$~\kms, comparable to that measured in Barnard's Loop \citep{Reynolds1979ApJ...229..942R}. Using the solution of \citet{Weaver1977ApJ...218..377W} the bubble expansion age can be estimated to be $\sim 0.7$~Myr. 
Given its redshifted velocity this blister is expanding away from us, downstream to the spiral arm, similar to what is seen in nearby galaxies \citep{Watkins2023ApJ...944L..24W}. 
With this small estimated age but accumulation of molecular material, is likely this bubble originated from supernovae feedback (as in the case of Barnard's Loop), and the association of cataloged OB stars with the rim of the GCL is suggestive of propagated star formation following shell compression driven in a past supernova event. However, it is unlikely to be associated with the very young (1~Myr) cluster currently at GCL-1.

\section{Conclusions}
\label{sec:conclusion}
We present a new SDSS-V LVM optical IFU map covering 1.5 $\times$ 3~deg in the direction of the Galactic Center, covering the so-called `Galactic Center' Lobe (GCL). We confirm recent results that identify this as a foreground object \citep{Nagoshi2019PASJ...71...80N, Tsuboi2020PASJ...72L..10T, Hurley-Walker2024ApJ...969...42H, Anderson2024ApJ...969...43A}, and not associated with the Galactic Center.  To this end, we suggest the acronym GCL be repurposed to instead abbreviate the name `Greatly Confused Loop'.  

In our work, we leverage new optical emission line constraints on the GCL to provide concrete answers to where, and what, the GCL is. 

\begin{itemize}
    \item Using \siii$\lambda$9532 emission line maps, which suffers the least extinction, we identify ionized gas emission throughout a full, closed outer loop, and further note that this high ionization ion does does not fill the bubble interior. There is a very good morphological correspondence between  the diffuse radio continuum emission, and 8.0$\mu$m PAH dust emission.
    \item All optical line diagnostics that we consider are consistent with photoionization. The lobe is bright in \ha, \siii, \sii\ and \nii. While detected in integrated spectra across the GCL, \oiii\ is not strongly detected in resolved maps. 
    \item Using Pa9/\ha\ and \ha/\hb\ we constrain R$_V$=3.6 and infer a dust reddening of E(B-V)=1.0 to 1.5~mag.  Combined with 3D dust maps, this directly constrains the distance to the GCL to $\sim$2~kpc.
    \item We estimate \te\nii=7000-8000~K based on strongline ratios, when compared to a sample of Galactic \hii\ regions, which implies slightly super-solar metallicity, and is strongly discrepant from the radio recombination line \te=3.500~K. Based on the Galactic radial \te\ gradient, this implies a distance of less than 4~kpc.
    \item \nii\ line kinematics show a uniform velocity structure across the GCL, further confirming that the entire bubble is one associated structure.
\end{itemize}

Given the $\sim$2~kpc distance of this photoionized structure, the size and emission line morphology is strongly reminiscent of what is seen in the nearby Barnard's Loop. It may indicate a similar origin, of a past supernova event clearing a cavity 35~pc in diameter, which is currently the site of new OB star formation and enables radiation to propagate through the cavity interior to sustain the photoionization of the shell.  The young blister-like ($\sim$1~Myr) cluster associated with GCL-1/RCW~138 may play a significant role in the photoionization of the GCL structure. 

Given the available multi-wavelength views of this region, we have shown that the specific dust extinction structure along this line of sight results in the ionized gas unveiling most accurately the closed, bubble-like morphology of the GCL, as everything more distant is completely extincted. Mid-IR PAH features and radio continuum emission are blended, which has obstructed the identification of this feature in past work. Parallels with Galactic Center X-ray features, and the base of the Fermi bubble, must be largely coincidental.

\begin{acknowledgements}
We thank the referee for their careful reading, which helped to improve the clarity and quality of this work.
Many thanks to Ian Heywood for providing the MeerKAT data over a larger field of view. 
KK gratefully acknowledges funding from the Deutsche Forschungsgemeinschaft (DFG, German Research Foundation) in the form of an Emmy Noether Research Group (grant number KR4598/2-1, PI Kreckel) and the European Research Council’s starting grant ERC StG-101077573 (“ISM-METALS"). 
KK and SK acknowledge funding by the Deutsche Forschungsgemeinschaft (DFG, German Research Foundation) – Project number 558818801.  
OE acknowledges funding from the Deutsche Forschungsgemeinschaft (DFG, German Research Foundation) -- project-ID 541068876. 
J.G.F-T gratefully acknowledges the grants support provided by ANID Fondecyt Postdoc No. 3230001 (Sponsoring researcher), the Joint Committee ESO-Government of Chile under the agreement 2023 ORP 062/2023, and the support of the Doctoral Program in Artificial Intelligence, DISC-UCN. JEM-D and R de J. Zermeño  thank the support by SECIHTI CBF-2025-I-2048 project ``Resolviendo la Física Interna de las Galaxias: De las Escalas Locales a la Estructura Global con el SDSS-V Local Volume Mapper'' (PI: Méndez-Delgado). JEM-D and R de J. Zermeño thank the support by UNAM/DGAPA/PAPIIT/IA103326 project ``DESIRED (DEep Spectra of Ionized Regions Database): de las emisiones más sutiles a la física fundamental del universo’’ (PI: Méndez-Delgado).
SFS acknowledges the support by CBF-2025-I-236 project granted by the Secretaría de Ciencia, Humanidades, Tecnología e Innovación (SECIHTI) of
the Mexican Federal Government, and the PID2022-136598NB-C31 (ESTALLIDOS) grant by the Spanish Ministery of Science and Innovation (MCINN).
A.Z.L.A. gratefully acknowledges the support provided by the Post-
doctoral Program (POSDOC) of UNAM (Universidad Nacional
Autónoma de México).

Funding for the Sloan Digital Sky Survey V has been provided by the Alfred P. Sloan Foundation, the Heising-Simons Foundation, the National Science Foundation, and the Participating Institutions. SDSS acknowledges support and resources from the Center for High-Performance Computing at the University of Utah. SDSS telescopes are located at Apache Point Observatory, funded by the Astrophysical Research Consortium and operated by New Mexico State University, and at Las Campanas Observatory, operated by the Carnegie Institution for Science. The SDSS web site is \url{www.sdss.org}. 
SDSS is managed by the Astrophysical Research Consortium for the Participating Institutions of the SDSS Collaboration, including the Carnegie Institution for Science, Chilean National Time Allocation Committee (CNTAC) ratified researchers, Caltech, the Gotham Participation Group, Harvard University, Heidelberg University, The Flatiron Institute, The Johns Hopkins University, L'Ecole polytechnique f\'{e}d\'{e}rale de Lausanne (EPFL), Leibniz-Institut f\"{u}r Astrophysik Potsdam (AIP), Max-Planck-Institut f\"{u}r Astronomie (MPIA Heidelberg), Max-Planck-Institut f\"{u}r Extraterrestrische Physik (MPE), Nanjing University, National Astronomical Observatories of China (NAOC), New Mexico State University, The Ohio State University, Pennsylvania State University, Smithsonian Astrophysical Observatory, Space Telescope Science Institute (STScI), the Stellar Astrophysics Participation Group, Universidad Nacional Aut\'{o}noma de M\'{e}xico, University of Arizona, University of Colorado Boulder, University of Illinois at Urbana-Champaign, University of Toronto, University of Utah, University of Virginia, Yale University, and Yunnan University. 
This work made use of the Cube Analysis and Rendering Tool for Astronomy \citep[{\tt CARTA};][]{2021zndo...3377984C}, % 2021ascl.soft03031C , angus_comrie_2018_3377984
and a number of python packages, namely the main \textsc{astropy} package \citep{Astropy+2013, Astropy+2018, Astropy+2022}, \textsc{numpy} \citep{Harris+2020}, \textsc{matplotlib} \citep{Hunter+2007} and \textsc{Spectral-Cube}.\footnote{\url{https://spectral-cube.readthedocs.io/en/latest/}}

\end{acknowledgements}

%-------------------------------------------------------------------
% \bibliographystyle{aa}
\bibliographystyle{aa_url}

\clearpage

\begin{appendix} %First online appendix

\section{Further details on the low Galactic latitude HII regions}
\label{app:blow}
\begin{figure*}
    \centering
    \includegraphics[width=0.95\linewidth]{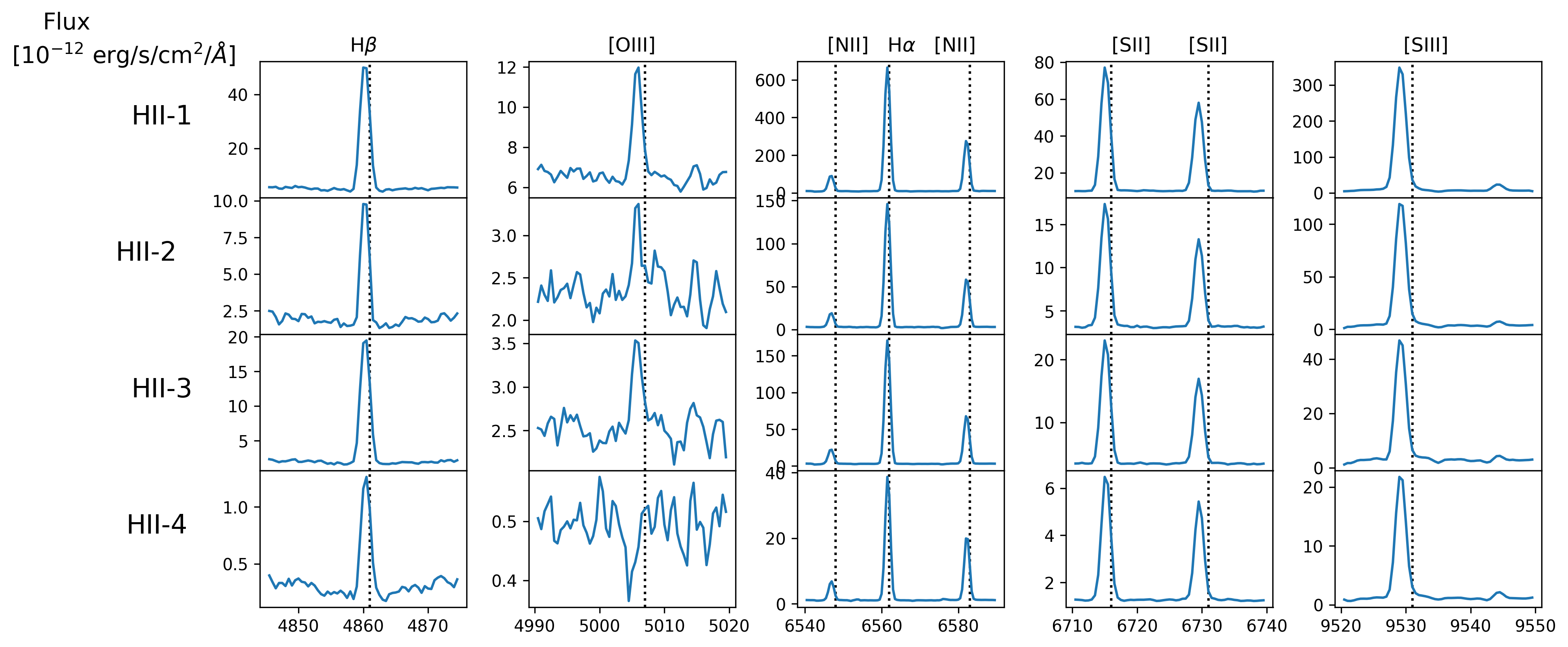}
    \caption{Key spectral features for all four b$<$0$^{\circ}$ \hii\ regions labeled in Figure \ref{fig:subregions}. Line centers (assuming v$_{LSR}$=0~\kms) are marked with vertical dotted lines.}
    \label{fig:blow_spec}
\end{figure*}

In Figure \ref{fig:blow_spec} we show spectral features of key emission lines, as detected in the integrated spectra of four \hii\ regions at b$<$0$^{\circ}$ (see Figure \ref{fig:subregions}). 

A limited amount of information is available for each of these in the literature, however they are generally thought to be foreground to the Galactic Center region. Cross-identification of these regions with existing catalogs and tentative distance estimates for each are provided. All have been confirmed to be \hii\ regions based on their radio recombination line emission \citep{Anderson2014ApJS..212....1A, Anderson2015ApJS..221...26A}, and are referenced in the The WISE Catalog of Galactic HII Regions.

\textbf{HII-1}, which corresponds to Sh 2-16, has an estimated distance of  2.526 $\pm$ 0.149 kpc based on the parallax distance to Gaia DR3  4057274876643924096 (LS 4381) of the apparent ionizing source. Its WISE identification is G359.756-00.351.

\textbf{HII-2}, which corresponds to Sh 2-20 and RCW 141, has an estimated distance of 3.909 $\pm$ 0.291 kpc \citep{Monteiro2019MNRAS.487.2385M}. Its WISE identification is G000.320-00.215.

\textbf{HII-3}, which corresponds to RCW 139, has an estimated distance of 2.41 $\pm$ 0.42 kpc based on the Gaia DR3 4057473097944289920 parallax of 0.4148 $\pm$ 0.0869 mas for a possible ionizing source. This is consistent with an older distance estimate of 1.7 $\pm$ 0.4 kpc \citep{Avedisova1989BAICz..40...42A}. 
Its WISE identification is G000.120-00.556.

\textbf{HII-4}, which corresponds to Sh 2-21 and RCW 142, has an estimated distance of  2.0 $\pm$ 0.4 kpc \citep{Schenck2011AJ....142...94S}. Its WISE identification is G000.573-00.855.

These are at similar (or slightly larger) distance compared to the GCL, however distinctly different kinematics (Figure \ref{fig:NII_vel}, suggesting that they are not in close physical proximity to the GCL. 

\end{appendix}

% \label{LastPage}
\end{document}